# EXPLAINABLE AI FOR TAILORED ELECTRICITY CONSUMPTION FEEDBACK – AN EXPERIMENTAL EVALUATION OF VISUALIZATIONS

*Research Paper*

Jacqueline Wastensteiner, University of Bamberg, Bamberg, Germany, jacqueline.wastensteiner@web.de

Tobias M. Weiss, University of Bamberg, Bamberg, Germany, tobias.m.weiss@gmx.de

Felix Haag, University of Bamberg, Bamberg, Germany, felix.haag@uni-bamberg.de

Konstantin Hopf, University of Bamberg, Bamberg, Germany, konstantin.hopf@uni-bamberg.de

## Abstract

*Machine learning (ML) methods can effectively analyse data, recognize patterns in them, and make high-quality predictions. Good predictions usually come along with "black-box" models that are unable to present the detected patterns in a human-readable way. Technical developments recently led to eXplainable Artificial Intelligence (XAI) techniques that aim to open such black-boxes and enable humans to gain new insights from detected patterns. We investigated the application of XAI in an area where specific insights can have a significant effect on consumer behaviour, namely electricity use. Knowing that specific feedback on individuals' electricity consumption triggers resource conservation, we created five visualizations with ML and XAI methods from electricity consumption time series for highly personalized feedback, considering existing domain-specific design knowledge. Our experimental evaluation with 152 participants showed that humans can assimilate the pattern displayed by XAI visualizations, but such visualizations should follow known visualization patterns to be well-understood by users.*

## 1 Motivation

Many outstanding applications of machine learning (ML)—a core technology of artificial intelligence (AI)— documented in the literature focus on their superiority in making predictions about unseen data or future events. Cancer detection from radiological images (McKinney et al., 2020) and fraud detection (Abbasi et al., 2012) are often cited examples for tasks in which ML reaches human levels or partly outperforms humans. Such applications relate to the use of *AI to automate* tasks in a wide range of industries (Coombs et al., 2020). We owe the performance of these AI applications to ML models that are becoming increasingly complex and difficult for humans to understand. Such ML models are often *black-boxes,* which come at the price of low interpretability (Dourish, 2016; Faraj et al., 2018). The opposite of these models are transparent ones, having lower capabilities to generalize from data (Barredo Arrieta et al., 2020). Motivated by this tension, a recent field of research in the area of ML has put forth eXplainable AI (XAI) approaches that make complex black-box models interpretable to humans, without lowering their predictive power (Miller, 2019). These approaches are promising, not only for applications where it is necessary to make algorithmic judgement interpretable to humans (e.g., for legal or ethical decisions), but also for applications where AI is employed to provide more insights to humans—uncover patterns in data, not only making predictions—enabling more informed human decisions. Therefore, XAI supports the use of AI to *augment work* (Grønsund and Aanestad, 2020; Raisch and Krakowski, 2020) instead of replacing humans (Frey and Osborne, 2017).

Current XAI approaches, however, are criticized for (i) focusing too heavily on technical aspects or data perspectives of developers and (ii) including few aspects of social sciences and human-computer interaction (Abdul et al., 2018; Miller, 2019). Existing XAI studies have also often focused on content, less on the interface design of explanations (Cheng et al., 2019). Literature also points to a lack of XAI user studies (Adadi and Berrada, 2018; Nourani et al., 2019). Similarly, recent calls from information systems research motivate empirical studies on the application of AI in organizations, not only to automate but to augment human labour (Coombs et al., 2020; Lyytinen et al., 2020; Rai et al., 2019).

Time series are common data structures that ML and XAI methods can process. Time is an important dimension of data analysis and time series data becomes more present, as digitization increases the proliferation of sensors and smart devices, which capture more data with timestamps. An area that could particularly benefit from uncovering and visualizing hidden patterns in time series data is residential energy consumption. Behavioural research demonstrates that specific feedback on consumers' energy consumption—tailored to individuals—leads to sustainable behaviour and, thus, can trigger significant energy savings (Brülisauer et al., 2020; Tiefenbeck et al., 2016). Deployed on a large scale, behavioural feedback interventions can play an essential role in lowering the energy demand, thus reducing the human carbon footprint. Although increasing amounts of data is available in the residential energy context (e.g., because of smart metering infrastructures), helpful behavioural recommendations are hard to extract from the data on a large scale. Advanced data processing and modelling techniques are therefore warranted to make undesired human behaviour salient, and to guide people towards better action. We believe that XAI can be of reasonable help in this regard and selected this case study for our research project. The context of this study is also well suited to put forth XAI visualisations, because plenty of time series data is available that contains complex patterns, which may be not easily to recognize by humans but for ML. Thus, we examine the following research question:

> *How well do XAI visualizations of electricity consumption time series data, created based on design knowledge from XAI and feedback research, perform in terms of comprehensibility of humans and user preferences?*

We created five XAI visualizations based on the current technological state as well as design knowledge from the XAI and feedback literature. In a user experiment with 152 participants, we evaluated these visualizations in isolation, using reading and memorization tasks, and in comparison, using a conjoint experiment. Our results show that XAI can provide insights into electricity consumption time series data that can be assimilated by humans. We also found that standard XAI visualizations should be adjusted to foster comprehensibility by humans. These results underline the need for further investigating XAI-based human-AI interfaces and tailored consumption feedback, as we outline in our discussion.

This paper proceeds with a review of the recent literature around XAI and automated feedback on residential electricity use. Thereafter, we describe our research approach, our case selection, and the design and implementation of XAI visualizations. Section 5 describes our experimental evaluation and our findings. We finish this paper with a discussion and formulate implications for future research.

## 2 Related work

The discourse on XAI technology takes place primarily in the field of computer science, where it has led to advances in the technological basis. Nevertheless, it would benefit from social science research (Miller, 2019) and business perspectives (Satell and Sutton, 2019). This is a type of contribution that lies at the core of the information systems research tradition, because this field purses a sociotechnical perspective (Sarker et al., 2019). Lyytinen et al. (2020) and Rai et al. (2019) underline the need for such research to better understand the successful integration of AI in workplaces.

## 2.1 XAI in information systems research

So far, information systems research has conceptualized the possibilities of XAI to enable personalized explanations of ML models (Schneider and Handali, 2019) and the compliance with recommendations that stem from AI (Kühl et al., 2019). Wanner et al. (2020) provide a literature review and outline a plan for a user study to investigate the willingness of users to dispense the accuracy of model prediction in favour of better explanations. Our work adds to this (so far conceptual) research an empirical investigation on the application of AI in the context of energy feedback. Thereby, we draw on the literature from XAI and feedback on energy consumption. We briefly summarize both areas below.

## 2.2 XAI technology

XAI is a very active field of research and technological development. This becomes visible in several comprehensive literature review articles on that topic, which provide taxonomies of current XAI approaches (e.g., Adadi and Berrada, 2018; Anjomshoae et al., 2019; Barredo Arrieta et al., 2020). Explanatory methods can be classified according to several criteria, namely their compatibility (model-specific vs. model-agnostic), the degree of interpretability (local vs. global), and whether ML models are directly interpretable (intrinsic) or require methods that analyse ML models after training (post-hoc). We concentrate on model-agnostic methods, which are mostly applied post-hoc and are pluggable on any ML model, which makes them independent of a particular class of ML algorithms. Within these group of model-agnostic methods, we focus on feature attribution methods that estimate the impact of features on predictions on a local level (i.e., the impact of each feature on each individual predicted instance). In the case of time-series prediction models, this capability allows to estimate the contribution of individual time periods for a given outcome. In this category, we focus on two methods that recent works (Slack et al., 2020) perceive as very relevant:

- *Deep Shapley Additive exPlanations (SHAP)* was introduced by Lundberg and Lee (2017) and uses concepts from game theory for the predictor variable importance estimation.
- *Local Interpretable Model-Agnostic Explanations (LIME)* introduces variations in the dataset (i.e., perturbation) and estimates how these affect the predictions of the black-box model by using a human-interpretable model, like linear regression (Ribeiro et al., 2016).

Research has carried out comparisons of these methods (Alvarez Melis and Jaakkola, 2018; Schlegel et al., 2019) using datasets from several domains. We have built upon these procedures and evaluated both XAI methods in our study.

Literature from the area of human computer interaction points to the importance of user studies to evaluate XAI visualizations (Abdul et al., 2018) and suggests to assess these visualizations with respective metrics on the comprehensibility and user preferences. Current XAI visualizations are, for example, criticized for being complex, target primarily ML-experts, and neglect a user perspective that would foster the understanding of the visualizations (Abdul et al., 2020; Kaur et al., 2020). Mohseni et al. (2020) develop a framework with design guidelines and evaluation methods to support the iterative design and evaluation loop of XAI visualizations.

## 2.3 Automated consumer feedback on electricity consumption

Feedback has received much attention in research and practice, because such behavioural intervention helps humans to overcome biases in their decision-making, thus it has the power to change human behaviour for the good (Allcott and Mullainathan, 2010). Feedback can lead to pro-ecologic behaviour (Klöckner, 2013) and can reduce energy use in the residential sector (Fischer, 2008; Karjalainen, 2011; Lu et al., 2016; Weiss et al., 2016) at comparable low cost (Benartzi et al., 2017). Behavioural research demonstrates that specific feedback to consumers, tailored to individuals, can lead to significant energy savings (Brülisauer et al., 2020; Tiefenbeck et al., 2016).

A major obstacle in realizing such tailored feedback in practice lies in missing data when generating personalized messages or visualizations on scale (Hopf, 2019, p. 147; Tiefenbeck, 2017). Collecting such data in surveys or with energy audits is costly. To overcome this problem, research has analysed electricity consumption time series data of private households with ML to extract the necessary information. One literature branch—the topic of non-intrusive appliances load monitoring—analyses consumption data of high frequencies (usually more than one measurement per minute) with the goal to detect single appliances (Hart, 1992; Zeifman and Roth, 2011), but such fine-grained data is usually not available in many households. Another branch of literature develops ML methods to detect more general household characteristics of residential households (Albert and Rajagopal, 2013; Beckel et al., 2014; Hopf, 2019; Hopf et al., 2018; Weigert et al., 2020). These approaches provide more viable aid to carry out feedback campaigns to many energy consumers. Results show, for example, that households with a single occupant can be identified with up to 81% accuracy, the type of the cooking facility with up to 87%, and certain heating systems with up to 85%. Models with high predictive performance in these works belong to the category of black-box models. Although extracted information about household characteristics is helpful to make feedback more specific, energy experts still must formulate energy saving recommendations based on predicted data. XAI has a high potential to overcome this drawback.

## 3 Research approach

Our study developed and evaluated XAI visualizations. Based on these artefacts, our objective was to generalize experiences that contribute to the current debate on how to create effective XAI visualizations. Our research approach followed the guidelines of design science research in information systems (Hevner et al., 2004; Peffers et al., 2007). More precisely, we took up the Ivari's (2015) second strategy to conduct design science research, that solves a specific problem (i.e., tailored feedback based on XAI) by building concrete IT artefacts in a specific context. From that we distil knowledge to address a class of problem (i.e., human-understandable visualizations of patterns in time-series data).

Our design and evaluation efforts draw on two research areas, each of which brings substantial literature: We combined a technical perspective (i.e., XAI) and a domain perspective (i.e., feedback on electricity consumption) while pursuing our research, as we illustrate in Figure 1. We describe the first step (case selection and problem definition) and the second step (requirement elicitation and definition of design features) in this section, the technical implementation and their experimental evaluation in the following.

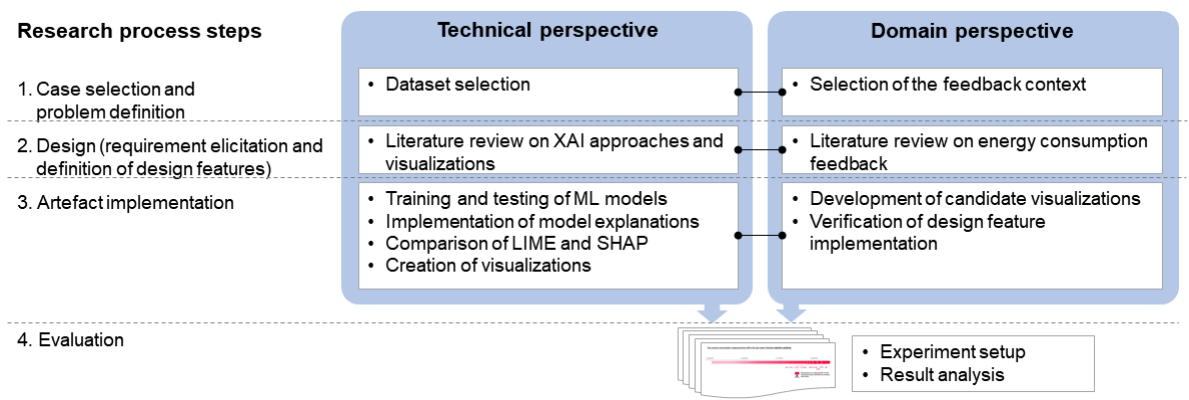

*Figure 1.     Research approach.*

### 3.1 Case selection and problem definition

ML applications require a sufficient amount of training data that consists of—in the case of predictive models—several predictor variables and ground truth data on the variable that should be predicted. Earlier works that analysed electricity consumption time series data (15-min or 30-min smart meter data) with ML for predicting household characteristics to support consumption feedback used datasets from North America (Albert and Rajagopal, 2013), Ireland (Beckel et al., 2014; Wang et al., 2018) and Switzerland (Hopf, 2019; Hopf et al., 2018). The largest dataset, which is also publicly available, stems from a smart meter trial from the Commission for Energy Regulation (2011) in Ireland and covers 30-minute smart meter electricity consumption data on 76 weeks (July 2009 – December 2010) and survey data for 4,232 households. The dataset also contains information on household characteristics ("ground truth data"). We selected this dataset for our study because it was the largest available dataset.

We reviewed the survey data and selected those variables on household characteristics that (i) are related to energy-intense activities, (ii) could potentially help to develop XAI electricity consumption feedback, and (iii) could be detected with a comparably good predictive performance in earlier household characteristics prediction studies. We thus selected: Electric cooking (yes, no), Presence at home during typical days (yes, no), and Electric water heating (yes, no). For each of the household characteristics, we trained ML models that predicted the respective variable. We applied XAI to visualize the times of electricity use that the ML algorithm detected as relevant, to generate informative visualizations for electricity consumption feedback. Details on the implementations and performance results follow in Section 4.

### 3.2 Design of XAI visualizations for feedback on electricity consumption

We conducted a comprehensive literature review in which we identified 8 requirement categories for XAI visualizations and 17 requirement categories for electricity consumption feedback (details on this review and the detailed list of design requirements and features are listed in the Appendix). Based on this design knowledge, we developed five basic XAI visualizations. Each visualization can fulfil the design requirements to a certain degree. Figure 2 shows an example of each type of visualization.

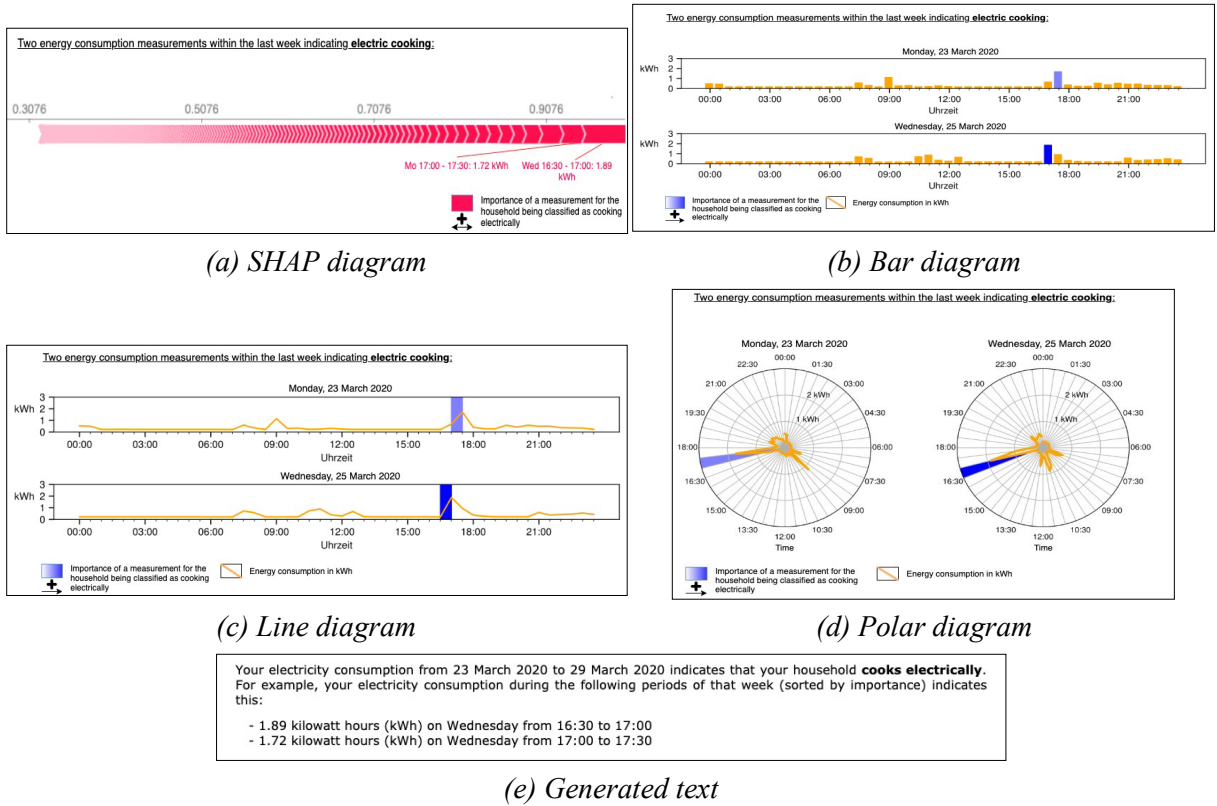

*(a) SHAP diagram*     *(b) Bar diagram*

*(c) Line diagram*     *(d) Polar diagram*

*(e) Generated text*

*Figure 2.       XAI visualizations evaluated in our experiment.*

The first diagram is a standard visualization of the SHAP approach. We included this to represent a state-of-the-art visualization of XAI. Then, we adopted four illustrations that follow recommendations of the energy feedback literature. A line diagram and a bar diagram, which are the most frequent visualizations of electricity consumption feedback (Herrmann et al., 2018). Both tie in with the natural analogy of taking electricity consumption data from left to right using a timeline. We also considered a polar diagram that links to a clock analogy where 24 hours of consumption data are displayed in a circle. Although users seem to perceive the line and bar diagram more positively and understand them better than the polar diagram (Flora and Banerjee, 2014), we wanted to evaluate to what extent the additional information from XAI on a clock analogy is understood by users. All diagrams contained a highlight in blue that indicated the period which was particularly relevant for the ML model decision to classify the household as the respective class (e.g., electric cooking). Finally, we considered a basic text description of the most relevant information from the ML models as a form of non-visualization.

## 4 Technical implementation

Our technical implementation[1] that generated the electricity feedback artefacts consisted of two steps, as Figure 3 illustrates. In Step A, we created a ML prediction model that was trained to predict an energy consumption related variable for each household. We are not primarily interested in the predictions of this model, rather in the patterns that this model detects in the electricity consumption data. The analyses we did were to verify that our implementation follows the current state-of-the-art in ML modelling. Step B then applied XAI methods to extract and visualize times of electricity use that the ML model found relevant. We compared the two XAI methods and selected the most suitable one. The description of our technical implementation focuses on essential aspects to understand the generated feedback element artefacts due to the focus of this paper and the limited space available.

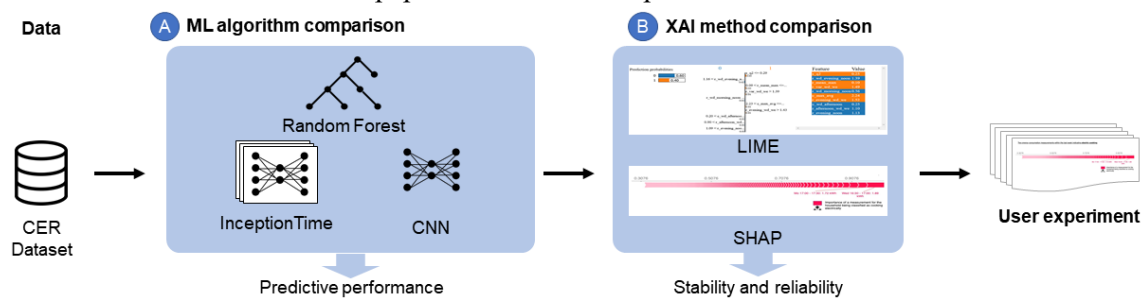

*Figure 3.   Overview technical implementation and the comparisons of the ML algorithms (A) and the XAI methods (B).*

### 4.1 ML model implementation and comparison

We considered three ML algorithms for the time series classification task. First, *Random Forest* (Breiman, 2001), an ensemble learner that combines multiple uncorrelated decision trees to obtain a well performing prediction in many real-world applications (Fernández-Delgado et al., 2014). Second, *convolutional neural network* (CNN), an approach from the field of deep learning. Previous studies found that CNN and Random Forest could detect household characteristics from electricity consumption smart meter data with good performance (Hopf, 2019; Wang et al., 2018), Third, the *InceptionTime* classifier (Ismail Fawaz et al., 2020, 2019), which combines an ensemble of five CNNs in that it parallelizes the convolutional layers. Ismail Fawaz et al. (2019) demonstrate that their approach achieves higher stability and prediction accuracy on time series data than other state-of-the-art classifiers. As an input, CNN and InceptionTime took each week of electricity consumption time series data together with labels for the respective household. Both algorithms can directly process image representations of time series data. For Random Forest, we follow earlier studies and extracted 93 predictor variables from the time series to reduce the dimensionality (Beckel et al., 2014; Hopf et al., 2018).

We compared the three ML algorithms regarding their *predictive performance* for the three selected dependent variables and list the results together with statistics on the original data in Table 1. As performance metrics, we used accuracy (ACC), which is the percentage of correctly classified observations in the test sample, and the area under the receiver operating characteristic curve (AUC). Both metrics are well-known for ML model evaluation (Hastie et al., 2009). Whereas ACC is easy to interpret, its values are biased by the class distributions. Therefore, ACC results of different variables cannot be compared. AUC can be used as an unbiased estimate of the predictive performance (Fawcett, 2006). For the performance evaluation, we follow good practices of ML evaluation and apply 10-fold cross-validation (Hastie et al., 2009) with a random allocation of the samples to the ten folds.

---

[1] The source code of our implementation is available at https://gitlab.rz.uni-bamberg.de/eesys-public/household-classification-explainable-ai  for further use.

| Variable | Sample size | | Relative Freq. positive class | InceptionTime | | CNN | | Random Forest | |
|---|---|---|---|---|---|---|---|---|---|
| | Households | Num. weeks | | ACC | AUC | ACC | AUC | ACC | AUC |
| Electric cooking | 4,232 | 114,455 | 69.91% | **0.72** | **0.67** | 0.72 | 0.61 | 0.70 | **0.69** |
| Presence at home during the day | 1,310 | 27,949 | 56.95% | 0.78 | 0.73 | 0.76 | 0.66 | 0.77 | 0.74 |
| Electric water heating | 4,232 | 138,044 | 80.63% | 0.63 | 0.62 | 0.60 | 0.59 | 0.62 | 0.63 |

*Table 1.    Statistics and predictive performance of both ML algorithms for the three considered dependent variables.*

We took a conservative modelling approach and changed the standard parameters of the algorithms with only a few variations to avoid bad configuration at chance[2]. More extensive optimization of hyper-parameters can certainly improve our results. Thus, the performance results lie within those of earlier studies, which used the same data set for the predictions of the cooking facility and achieved ACC between 0.69 and 0.71 with different non-deep-learning ML algorithms (Beckel et al., 2014), and between 0.739 and 0.766 (Wang et al., 2018), using CNN-based approaches. Wang et al. (2018), for example, used hyper-parameter tuning to optimize their performance.

## 4.2   Implementation and comparison of the XAI methods

To extract human-comprehensible visualizations from *InceptionTime* (the best performing prediction model in our analysis), we applied the XAI approaches SHAP and LIME. Both methods estimate the importance of certain predictor variables on the level of individual observations. In our case, each approach estimated which time span was particularly (ir)relevant to classify a household as electric cooking or not electric cooking. We compare both methods according to their *faithfulness* and *stability*, as suggested by Alvares Melis and Jaakkola (2018) and describe the evaluation procedures below.

*Faithfulness*: "Interpretability methods should … generate meaningful explanations … [even in the case of] local perturbations of the input … adding minimal [amount of] noise to the input" (Alvarez Melis and Jaakkola, 2018, p. 7). To operationalize this criterion, we adopted Schlegel et al.'s (2019) approach and modified the time series input data, by blurring values of predictor variables that were identified by the XAI methods to be most relevant for the model[3]. When the predictor variables are truly relevant for the prediction, the outcome should change considerably with such a data modification. We measured the relative amount of prediction changes after having modified 50 randomly chosen households (see Table 2).

*Stability*: This approach measured the ability of an XAI method to determine similar predictor variables for similar classifications. In doing so, we exploited the property of the time series data and the presence of daily routines of humans like, e.g., cooking during the same times of the day. We measured how often the ML model considered the same time of the day on different days as important for the model. We computed this frequency using a random selection of 50 households and weeks (see Table 2).

Based on the empirical analyses, *we finally selected the variable electric cooking with the InceptionTime predictor and the SHAP explainer for our further study*. The reasons were that these models showed a comparable high predictive performance, and that the variable provided the most reasonable feedback (advice to energy users based on electric cooking allowed for more actionable insights than based on other variables). Furthermore, the stability and faithfulness for this variable were comparably high.

---

[2] Our instance of InceptionTime used the parameters: Max. kernel size: 40, Depth: 6, Num. kernels: 32, Batch size: 64, Use Bottleneck: true, Use residual: true. For numeric parameters, we tested three alternatives, for the binary parameters both values. We selected the best performing setting based on AUC on a 20% sample of the data. Calculations for InceptionTime ran on Python Keras 2.2.4. For Random Forest, ntree=100 was used. The computations ran on Python using scikit-learn 0.19.1.

[3] The replacement was the deviation of the consumption measurement from the average consumption of the household in the opposite direction, as a zero consumption or negative consumption could be recognized by the ML algorithm as a special case.

|  | Faithfulness (relative number of predicted changes) | | Stability (relative number of non-unique time stamps) | |
| --- | --- | --- | --- | --- |
|  | SHAP | LIME | SHAP | LIME |
| Presence | 0.24 | 0.22 | 0.32 | 0.28 |
| Water heating | 0.38 | 0.26 | 0.35 | 0.31 |
| Cooking | 0.3 | 0.24 | 0.42 | 0.35 |

*Table 2.     Comparison of the two XAI methods LIME and SHAP for the three selected household characteristics regarding the criteria faithfulness and stability.*

## 5   Experimental evaluation of the five visualizations

We conducted an experimental evaluation of the five obtained visualizations. The experiment was carried out as an online survey and had two phases: The first phase focused on the isolated evaluation of each visualization. We collected subjective (self-reported) and behavioural measures to evaluate the visualizations. In the second phase of the experiment, we used a choice-based conjoint to measure user preferences on the visualizations. Before the experiments started, we asked for sociodemographic variables. In total, the online experiment took 17:16 minutes on average (13:30 minutes standard deviation).

### 5.1   Sample description

We promoted the survey among students of our institution and used several online channels to attract participants outside of the university context. Our sample is balanced regarding the gender (51.32% female, 48.68% male, 0% diverse / not given), but it has a bias towards younger participants with higher education (82.9% are younger than 35; in the German population, only 36.7% are in this age category), likely because many participants were students from the university. However, the share of participants which were employed (not marginally employed) is 44.1% which is similar to the share of employed citizen and civil servants in Germany, which is 45.0% (DESTATIS, 2020, p. 39). Participants lived more frequently in rented homes (69.1%) than the population (48.9%), according to Eurostat (2020), but the number of people living in the households was similar to the German average ($r(3) = .96$, $p = .011$).

### 5.2   First phase: Isolated evaluation of the visualizations

The first phase of the experiment evaluated the comprehensibility of the five electricity feedback visualizations (see Figure 2). We first describe the experimental setup and then analyse the results.

#### 5.2.1   Experimental setup

We carried out four reading and memory tasks with the participants, each time with one randomly selected visualization out of the five that we generated. Each participant saw each visualization only once. Reading and memory tasks are common for evaluating XAI visualizations (Abdul et al., 2020). We instructed participants to study (and memorize) the energy feedback illustration and informed them that the illustration would not be shown when answering subsequent questions. In total, we collected eight variables in the first part of the experiment (see Table 3).

The memorization task measured their *objective understanding* (Abdul et al., 2020; Cheng et al., 2019). For that, we asked them to rate three statements regarding the visualization as correct or incorrect. The statements had comparable length (Yan and Tourangeau, 2008) and were randomly selected from three preformulated sets of statements to avoid memory effects in the series of tasks. Each of the three statement sets consisted of two correct and two incorrect statements. The sets covered the topics (1) electricity consumption at specific times, (2) the prediction made by machine learning, and (3) the model explanation. Participants could also select an "I don't know" alternative for each statement. After the memorization task, participants indicated their *mental effort* for completing the task (Paas et al., 2008) on a seven point Likert scale. Finally, they indicated their *subjective understanding* of the visualization (Cheng et al., 2019) in terms of a German school grade from "1" (best) to "6" (insufficient). All used survey instruments can be requested from the authors. In addition to the self-reported data, we measured the reading and completion time of the tasks to collect objective behavioural data. In the online system, backward navigation was disabled, i.e., after the participants have seen the visualization and accessed the page with the follow-up questions, they could no longer see the visualization. In this way we ensured that participants had to answer the questions from their memory.

### 5.2.2 Statistical analysis and results

We analysed the results of the first experiment with an ordinary least squares linear regression. For each evaluation metric that we collected during the first phase of our experiment, we estimated one model (see results in Table 4). The models follow the specification $Y_i = \beta_0 + \beta_1 * VISUAL_i + \beta_2 * Age_i + \beta_3 * Edu_i + \epsilon_i$. $Y$ was the dependent variable (see Table 3, variables 1-6) that we collected in one of the four memorization experiments that each participant completed. $VISUAL$ is a categorical variable related to the visualization that was displayed to the participant. We used a dummy-encoding to represent

| | Variable | Description | Values | Mean (Std. dev.) or frequency |
|---|---|---|---|---|
| 1 | ReadingTime | The time (in seconds) each participant spent on reading the visualization. We compute the natural logarithm from the measurements to reduce the positive skew of the empirical distribution. | $\mathbb{R}^+$ | 3.47 (0.69) |
| 2 | AnswerTime | The time (in seconds) each participant spent on answering the questions for the visualization. We also computed the natural logarithm. | $\mathbb{R}^+$ | 3.13 (0.63) |
| 3 | MemTaskRight | Number of correct answers in the memorization task. | $[0,3] \in \mathbb{N}$ | 2.05 (0.89) |
| 4 | MemTaskDontKnow | Number of "I don't know" answers in the memorization task. | $[0,3] \in \mathbb{N}$ | |
| 5 | MentalEffort | Self-reported mental effort during completing the recall experiments, on a seven-point Likert scale. | $[1,7] \in \mathbb{N}$ | 3.67 (1.38) |
| 6 | SchoolGrade | Self-reported school grade participants estimated on their result on the recall tasks. | $[1,6] \in \mathbb{N}$ | 3.91 (1.19) |
| 7 | Age | The age reported by survey participants. | $\mathbb{R}^+$ | 30.2 (10.6) |
| 8 | Education | This binary variable state whether the study participant gained a general qualification for university entrance in Germany or a lower education degree. | high school diploma | 0.875 (n=133) |
| | | | other | 0.125 (n=19) |

*Table 3.      Overview of variables raised in the experiment and used in the statistical analysis.*

| | ReadingTime | AnswerTime | MemTaskRight | MemTaskDontKnow | MentalEffort | SchoolGrade |
|---|---|---|---|---|---|---|
| (Intercept) | 3.51*** | 2.80*** | 1.72*** | 0.60*** | 3.31*** | 3.37*** |
| | (0.15) | (0.14) | (0.18) | (0.16) | (0.26) | (0.25) |
| VisualLine | 0.07 | 0.08 | 0.34** | -0.26* | 0.44* | -0.37* |
| | (0.09) | (0.08) | (0.12) | (0.10) | (0.17) | (0.15) |
| VisualBar | -0.02 | 0.08 | 0.20 | -0.13 | 0.32 | -0.20 |
| | (0.09) | (0.08) | (0.12) | (0.11) | (0.18) | (0.16) |
| VisualPolar | 0.18 | 0.11 | 0.06 | -0.05 | 0.29 | -0.04 |
| | (0.10) | (0.09) | (0.13) | (0.12) | (0.19) | (0.17) |
| VisualText | -0.14 | 0.15* | 0.19 | -0.25* | 0.54*** | -0.36* |
| | (0.08) | (0.07) | (0.11) | (0.10) | (0.16) | (0.14) |
| Age | 0.01* | 0.02*** | 0.01* | -0.00 | 0.00 | 0.00 |
| | (0.00) | (0.00) | (0.00) | (0.00) | (0.01) | (0.00) |
| EduHIGH | -0.27** | -0.26** | -0.05 | 0.05 | 0.00 | -0.21 |
| | (0.10) | (0.09) | (0.11) | (0.10) | (0.15) | (0.15) |
| R^2 | 0.06 | 0.11 | 0.03 | 0.02 | 0.02 | 0.02 |
| Adj. R^2 | 0.05 | 0.10 | 0.02 | 0.01 | 0.01 | 0.01 |
| Num. obs. | 608 | 608 | 608 | 608 | 608 | 608 |

*Asterisks indicate statistical significance (\*\*\* $p < 0.001$; \*\* $p < 0.01$; \* $p < 0.05$), standard errors are in parentheses*

*Table 4.      Statistical evaluation of the first experimental phase.*

the five different visualizations and choose SHAP as the reference level, given that it is the state-of-the-art visualization from the chosen XAI approach. $Age$ is a numeric variable of the participants' age in years and $Edu$ is a dummy variable with the value 1 for high school diploma and 0 for a lower degree. We used robust standard errors (White, 1980; Zeileis, 2004) and checked for homoscedasticity of the errors $\epsilon$.

In general, the reading and answer times of the different visualizations display little difference. Considering the other metrics, the SHAP illustration does not perform well. All other visualizations lead to higher task performance (number of right answers and number of responses with "I don't know"). For the line chart and the text display, the differences are statistically significant. This performance is confirmed by the subjective ratings with the school grade (lower numbers means better results). Interestingly, the mental effort for the SHAP illustration was reported lower than for the others, likely because of more "I don't know" answers.

## 5.3 Second phase: Choice-based conjoint

The conjoint experiment allowed us to estimate user preferences regarding the visualizations. The method originates from marketing research and is increasingly used in information systems research, particularly to evaluate the design of information systems, as Naous & Legner (2017) found in their literature review. We follow Naous & Legner's (2017) framework of conducting conjoint experiments in that we conducted a choice-based conjoint (CBC) in which the study participants had to choose between two alternative feedback elements.

### 5.3.1 Experimental setup

Our main interest in this experiment phase was to find out which visualization the study participants preferred. Following recommendations for conducting conjoint experiments (Backhaus et al., 2015; Naous and Legner, 2017), we tried to make the choice options more realistic and at the same time implement further design requirements in the field of energy feedback. Specifically, we varied the visualizations with an additional explanatory text, energy saving tips and a chatbot frame. The energy saving tip was included because earlier literature from energy feedback underlined the relevance of such feedback devices. We considered two variants of tips (Vasseur et al., 2019): A curtailment tip (CMT) that suggests thinking about a repetitive, habitual change to reduce its electricity consumption, and an efficiency tip (ET) that recommends lowering the household's electricity demand by making a one-time investment. In total, the variations of the presented choices varied in four stimuli (visualization types, existence of explanatory text, type of energy saving tip, and chatbot). We used a full profile approach in which all possible combinations of the stimuli ($5 \cdot 2 \cdot 2 \cdot 2 = 40$ variants in total) were considered. From all possible combinations, five choice sets were created for each study participant. Each choice set contained two randomly drawn variants together with a non-option as a third choice. The none-option makes the choice experiment more realistic (Vermeulen et al., 2008), because forced choice situations are avoided (Backhaus et al., 2015, p. 181).

### 5.3.2 Statistical analysis and results

To evaluate the conjoint experiment, we estimate a logistic regression with maximum-likelihood method to model the choices. The model has the specification $P(Y_i = 1) = \left(1 + \exp(-x_i^\top \boldsymbol{\beta})\right)^{-1}$ with the linear predictor $x_i^\top \boldsymbol{\beta} = \beta_0 + \beta_1 * VISUAL_i + \beta_2 * TEXT_i + \beta_3 * CHAT_i + \beta_4 * TIPP_i + \beta_5 * NONE_i + \epsilon_i$. $Y$ is the dependent variable that indicates whether an option was selected. $VISUAL$ is a categorical variable with the visualization that was displayed to the participant. The existence of an explainable text was a separate characteristic and represented with the variable $TEXT$. Further characteristics are if the visualization was embedded in a chatbot environment ($CHAT$) and the type of energy saving tip was displayed ($TIPP$). As usual in conjoint analyses, model the stimuli variables with effect-encoding. Only the variable ($NONE$) that specifies the no-option is encoded as a dummy (Vermeulen et al., 2008).

Table 5 shows the estimated model (the column *Estimate* contains the log odds) details together with the odds ratios. Users preferred the Line and the Bar visualization, given that both have an odds ratio of 2.64 and 2.7 respectively, which means that the chance of selection for these visualizations is 2.64 (2.7) times more likely than for the others. The SHAP illustration, which is the response category in the regression analysis, must be computed by summarizing all other estimates, has only an odds ratio of 0.29, so participants strongly prefer the line or bar visualization instead of the SHAP illustration.

|  | Estimate | Std. Error | z-value | p-value | Odds ratio |
|---|---|---|---|---|---|
| VisualNo | -0.42 | (0.11) | -3.862 | <0.001 *** | 0.66 |
| VisualLine | 0.97 | (0.12) | 8.242 | <0.001 *** | **2.64** |
| VisualBar | 0.99 | (0.11) | 8.772 | <0.001 *** | **2.70** |
| VisualPolar | -0.32 | (0.11) | -2.943 | 0.003 ** | 0.73 |
| *VisualSHAP* |  |  |  |  | *0.29* |
| Text_No | -0.10 | (0.06) | -1.752 | 0.080 | 0.91 |
| Chatbot_No | 0.03 | (0.06) | 0.550 | 0.582 | 1.03 |
| Tip_CM | 0.18 | (0.06) | 3.179 | 0.015 ** | 1.20 |
| None | -1.66 | (0.10) | -16.630 | <0.001 *** | 0.19 |
| AIC | 2498.82 |  |  |  |  |
| BIC | 2544.56 |  |  | *Asterisks indicate statistical significance* |  |
| Log Likelihood | -1241.41 |  |  | *(*** $p < 0.001$; ** $p < 0.01$; * $p < 0.05$),* |  |
| Deviance | 2482.82 |  |  | *standard errors are in parentheses* |  |
| Num. obs. | 2247 |  |  |  |  |

*Table 5.     Logistic regression results of the conjoint analysis.*

## 6 Discussion and Research Implications

Our experimental evaluation led to two findings that we outline in Table 6. This section discusses them, names limitations, and formulates the implications as well as future research needs for the field of electricity consumption feedback and human-AI interfaces.

| **Finding** | **Implication for consumption feedback** | **Implication for human-AI interfaces** |
|---|---|---|
| XAI technology can be used to develop tailored electricity consumption feedback for end-users | - New class of feedback elements based on XAI that display novel patterns in the data<br>- Feedback can be more tailored to individuals | - XAI can be a support to realize augmented reality, where humans are supported by ML<br>- Novel visualizations highlight patterns in time series data |
| The SHAP visualization has not performed well in comparison to others (especially the line diagram) | - Integrate XAI elements into existing feedback elements | - (Re-)align the design of human-AI interfaces with known standards (e.g., time series visualization) |

*Table 6.     Overview to findings and implications from our study.*

### 6.1 Summary of the major findings

Our experiment provides two important findings: First, our study demonstrated that XAI technology can help to develop tailored electricity consumption feedback. Our experiments showed that users can assimilate novel insights from time series data with them. The artifacts created in this study realize many requirements of XAI or feedback visualizations from earlier research. Our study further demonstrated that XAI-based electricity consumption feedback can constitute a new class of feedback, which can also be transferred to other domains (e.g., heating, anticipatory driving). Second, the SHAP diagram, a state-or-the-art visualisation in XAI, did not perform well compared to the other tested visualizations. The line visualization, in particular, performed better in both phases of the experiment. We suppose that this is due to two reasons: a) Given the natural analogy that depicts time series data on a timeline from left to right, this illustration might be easier to comprehend by humans; b) it leverages already known elements can help non-expert users (i.e., without prior domain-knowledge) to make sense of unfamiliar visualizations (Lee et al., 2016). The text, generated by XAI, had a better comprehensibility by participants but was less preferred in the conjoint experiment. From the second finding we conclude that results of XAI should be integrated into visualizations that follow known standards to foster receptivity by humans.

### 6.2 Limitations

Our study is one of the early investigations of XAI applications in information systems research and, to the best of our knowledge, the first application of state-of-the-art XAI technology in the area of residential electricity consumption to develop tailored feedback for consumers. Given this novelty and the broad scope of the study, we identified four limitations. First, a common problem in XAI evaluation is that ground truth for the explanations—obtained by ML—is missing. Gathering such data would be expensive, but this would significantly help to improve the approaches. Second, we could not clearly identify whether the variance in performance of the visualizations results from the visualization itself or if it results from the fact that detected pattern are not fully clear to the user. We had to make simplifications in our experiment, for example, we could not capture all potential user preferences (e.g., colour preferences, aesthetic design), and—due to the already long online survey—we have not controlled for graph literacy, which is recommended by Abdul et al. (2020). In addition, we did not control for energy literacy, while prior knowledge may have an impact on how users make sense of the presented visualizations (Herrmann et al., 2018; Quintal et al., 2016). Third, energy consumption related statements that we could generate from the available dataset (i.e., electric cooking) had only limited relevance in practice, because feedback on the cooking type, or activities related to cooking are hard to change behaviours. Future research could collect data on human activities that have more actionable impact on the consumption behaviour (e.g., standby consumption of appliances or old devices). Fourth, the experimental evaluation only considers SHAP based XAI visualizations. We focused on SHAP because it performed best in terms of stability and faithfulness for the case at hand. Nevertheless, future research could involve additional visualizations based on other XAI methods such as LIME.

### 6.3 Future research

Considering the two major findings and the three limitations of our study, we identify the following five areas for future research.

First, the *patterns that are detected by ML and visualized by XAI should be validated* with respect to their meaningfulness— separately to the visualization. Future studies could either investigate this with ground truth data, for example, collecting data on the true pattern of electric cooking in our case (e.g., with interviews, household surveys, or energy audits). Studies could also approach this using synthetic data where the pattern are known upfront, as for example Tonekaboni et al. (2020) did.

Second, the *visualization variants should be evaluated independent of the meaningfulness of the detected pattern*. Here, our experimental setting can be replicated using visualization of pattern that are known to be correct. This can reduce variance in the collected variables and should follow Abdul et al. (2020).

Third, the *efficacy of electricity consumption feedback should be validated in field trials* that measure the true conservation of resources. Future studies can, for example, use earlier studies from electricity (Allcott, 2011) and water consumption feedback (Tiefenbeck et al., 2016) as a blueprint to evaluate the novel XAI-based feedback visualizations.

Fourth, several other methods for time series data processing and XAI exist, which are steadily improved and novel ones suggested. Our research design can be *extended with alternative technical approaches*.

Fifth, further research could focus on feedback elements that *show what type of activity contributes how much to the overall electricity consumption*. The new XAI explanations could be embedded in interactive energy feedback displays that already depict the main energy consuming appliances in specific time-of-use frames (Costanza et al., 2012).

## 7 Conclusion

Our study evaluated five visualizations generated by current ML and XAI methods to give consumers feedback on their electricity consumption. We selected residential electricity consumption as our study context because reducing energy demand is a societal challenge. Yet, the energy consumption context is also an interesting study site from an information systems perspective, because extensive time series data is available, which contains complex patterns that may not be easy to recognize by humans. Given the recent calls for empirical research to get a better understanding on how to integrate AI in human workplaces (Lyytinen et al., 2020; Rai et al., 2019), and the importance of AI technology to support humans by augmenting reality rather than replacing humans by AI (Raisch and Krakowski, 2020), our study demonstrated the power of XAI methods in human-AI interface design and highlights areas of further research and development.

# 8 Appendix: Requirements and design features for XAI-based feedback on electricity consumption

We reviewed the related research fields to identify design requirements for XAI-based feedback on electricity consumption. As a starting point, we chose five XAI literature review articles (Abdul et al., 2018; Adadi and Berrada, 2018; Anjomshoae et al., 2019; Miller, 2019; Mohseni et al., 2020) and articles that summarize general feedback research (Cianci et al., 2010; Mumm and Mutlu, 2011; van Duijvenvoorde et al., 2008), pro-ecological behaviour (Klöckner, 2013), energy consumption feedback (Lu et al., 2016), and electricity feedback (Benartzi et al., 2017; Fischer, 2008; Karjalainen, 2011; Weiss et al., 2016). With these articles, we conducted a forward and backward search (we reviewed all references in the review articles and all citations of them in Google Scholar) in order to complete our picture on the topics. In total, we reviewed the metadata of 869 referenced articles in the XAI papers (376 in the electricity feedback papers) and 1,124 articles that cited these papers (2,870 for feedback). In this review, we selected papers that contained requirements for the design of novel feedback elements. In the end, we found ten additional articles for XAI visualizations and 16 additional articles for feedback regarding electricity use in addition to those that we used as the starting point of our review. We list the identified requirements and their realization in the visualizations in Table 7.

| Category | Requirement named in literature | XAI visualizations | | | | | Tips | | Chat-bot |
|---|---|---|---|---|---|---|---|---|---|
| | | LD | BD | PD | SHD | TX | CMT | ET | |
| *1. Requirements for XAI visualizations* | | | | | | | | | |
| 1.1. Information content | Why / why not explanations | ✓ | ✓ | ✓ | ✓ | ✓ | | | |
| | Display of few representative instances for why / why not explanations | O | O | O | O | O | | | |
| | Details about causal relations (selective explanations) | ✓ | ✓ | ✓ | ✓ | ✓ | | | |
| | Input variable information (value and relevance of the variables) | ✓ | ✓ | ✓ | ✓ | O | | | |
| | No display of accuracy information | ✓ | ✓ | ✓ | O | ✓ | | | |
| | No possibilities to modify the model in the case of high accuracy | ✓ | ✓ | ✓ | ✓ | ✓ | | | |
| 1.2. User interface | Combination of text and image elements | ✗ | ✗ | ✗ | ✗ | ✗ | | | |
| | Adequate degree of user interaction | ✓ | ✓ | ✓ | ✓ | ✓ | | | O |
| *2. Requirements for electricity consumption feedback* | | | | | | | | | |
| 2.1. Information content | Data source (actually measured el. consumption) | ✓ | ✓ | ✓ | ✓ | ✓ | O | O | |
| | Unit of measurement (kWh cost) | ✓ | ✓ | ✓ | ✓ | ✓ | ✗ | ✗ | |
| | Relation to time of usage | ✓ | ✓ | ✓ | ✓ | ✓ | ✗ | ✗ | |
| | Granularity related to activities | ✓ | ✓ | ✓ | ✓ | ✓ | ✓ | ✓ | |
| | Historical comparison with previous time units | ✗ | ✗ | ✗ | ✗ | ✗ | | | |
| | Descriptive, normative comparison with the average | O | O | O | O | O | ✓ | ✓ | |
| | Individualized energy saving tips | ✗ | ✗ | ✗ | ✗ | ✗ | ✓ | ✓ | |
| 2.2 Multimodal feedback | Combination of feedback types | O | O | O | O | ✗ | ✗ | ✗ | |
| 2.3. Specific formats | Bar diagram for historical comparison | O | ✓ | ✗ | ✗ | ✗ | | | |
| | Bar diagram for normative comparison | ✗ | ✓ | ✗ | ✗ | ✗ | ✗ | ✗ | |
| | Grading scale for injunctive normative comparison | ✗ | ✗ | ✗ | ✗ | ✗ | ✓ | ✓ | |
| 2.4. Colour usage | Use of diagram colours that are preferred by users (e.g., traffic light indicators) | O | O | O | O | ✗ | ✓ | ✓ | |
| | Colours that activate associations (e.g., red, green) | ✗ | ✗ | ✗ | ✗ | ✗ | ✓ | ✓ | |
| | Colours without associations (e.g., black, white) | ✓ | ✓ | ✓ | ✓ | ✓ | ✓ | ✓ | |
| | Text colours (black text on white background) | ✓ | ✓ | ✓ | ✓ | O | ✓ | ✓ | |
| 2.5. Interaction design | Evaluative feedback | ✗ | ✗ | ✗ | ✗ | ✗ | ✓ | ✓ | ✗ |
| | Interaction with virtual agent | ✗ | ✗ | ✗ | ✗ | ✗ | ✗ | ✗ | O |

*LD: line diagram, BD: bar diagram, PD: polar diagram, SHD: SHAP diagram, TX: text, CMT: curtailment tip, ET: efficiency tip*

Table 7. *Requirements for XAI and electricity consumption feedback; symbols indicate their realization ("✓" realized, "O" partly realized, "✗" not realized) to what extent our visualizations.*